\documentclass[conference]{IEEEtran}

\usepackage{amsmath,graphicx}

\usepackage[T1]{fontenc}
\usepackage{graphicx}
\usepackage{cite}
\usepackage{subfig}
\usepackage{xcolor}
\usepackage{subfig}
\usepackage{optidef}
\usepackage{amssymb}
\usepackage{amsmath}
\usepackage{array}
\usepackage{mathtools}
\usepackage{scalerel}
\usepackage{algorithm}
\usepackage{algorithmic}

\ifCLASSINFOpdf
\else
\fi
\begin{document}
\title{Deep Learning-based Phase Reconfiguration for Intelligent Reflecting Surfaces}

\author{\IEEEauthorblockN{{\"O}zgecan {\"O}zdo\u{g}an, \ Emil Bj{\"o}rnson,\\
		Department of Electrical Engineering (ISY), Link{\"o}ping University, Sweden}}

\maketitle

\begin{abstract}
Intelligent reflecting surfaces (IRSs), consisting of reconfigurable metamaterials, have recently attracted attention as a promising cost-effective technology that can bring new features to wireless communications.  These surfaces can be used to partially control the propagation environment and can potentially provide a power gain that is proportional to the square of the number of IRS elements when configured in a proper way.  However, the configuration of the local phase matrix at the IRSs can be quite a challenging task since they are purposely designed to not have any active components, therefore, they are not able to process any pilot signal. In addition, a large number of elements at the IRS may create a huge training overhead.   In this paper, we present a deep learning (DL) approach for phase reconfiguration at an IRS in order to learn and make use of the local propagation environment.  The proposed method uses the received pilot signals reflected through the IRS to train the deep feedforward network.  The performance of the proposed approach is evaluated and the numerical results are presented.
\end{abstract}

\IEEEpeerreviewmaketitle

\section{Introduction}

An intelligent reflecting surface (IRS), also known under the names reconfigurable intelligent surface \cite{Huang2018a} and software-controlled metasurface \cite{Liaskos2018a}, is a thin two-dimensional metasurface that is used to aid communications\cite{Wu2019a}.  According to the application of interest, an IRS has the ability to control and transform electromagnetic waves that are impinging on it. Recently, it has received a  massive attention from the academia and sometimes marketed as one of the key enabling technologies for the next generation wireless communication systems.

Bringing such a technology into reality requires to addrees many practical challenges. For instance, the proper configuration of an IRS critically depends on accurate channel state information (CSI).  However, there are two main issues that complicates the channel acquisition with IRS \cite{bjrnson2020}. First,  the IRS is not inherently equipped with transceiver chains. Therefore, it can not sense the pilot signals.  Besides,  introducing an IRS into an existing setup will increase the number of channel coefficients proportionally to the number of IRS elements.

In the literature, some deep learning (DL) solutions are discussed to tackle these problems \cite{Elbir2020survey}. In \cite{Elbir2020}, a supervised learning approach is presented where two identical convolutional neural networks (CNNs) are trained to estimate the direct and cascaded channels. In \cite{Huang2019}, a feedforward neural network is proposed to unveil the mapping between the measured user coordinates and the optimal phase matrix at the IRS that maximzes the targeted user's signal strength. Another approach is to equip the IRS with a small number of active elements with sensing capabilities. The data collected from the active elements are utilized during the training of deep neural networks (DNNs) in \cite{Taha2019, Jiang2020} and the underlying channel structure is exploited to learn the entire channel. There are also deep reinforcement learning based methods that aim to solve  the problem of joint optimization of IRS phases and transmit beamforming  assuming perfect CSI  \cite{Huang2020,Feng2020}.

In this paper, we propose a novel DL approach for phase-configuration in an IRS-assisted MIMO system. We design two DNNs that are fed by the received pilot signals to directly find the mapping between the pilot signals and the optimum phase matrix and downlink transmit beamforming vector, thereby bypassing the conventional intermediate step of estimating the channels, which is prone to error propagation. In the first DNN, we send full-length pilot sequences and compare our results with a conventional least-square (LS) estimator based scheme. In the second method, our goal is to reduce the pilot overhead. We train the DNN with  shorter pilot sequences and predict the optimum phases and beamforming vector at the online stage.

\textit{Notation}: Lower and upper case boldface letters are used for vectors and matrices, respectively.  The transpose and Hermitian transpose of a matrix $\mathbf{A}$ are written as $\mathbf{A}^T$ and $\mathbf{A}^H$, respectively. The superscript $(.)^*$ denotes the complex conjugate. The operation $\mathbf{A} = \mathrm{diag}\left( \mathbf{a}\right)$  with  $ \mathbf{a} \in \mathbb{C}^{N\times 1}$ returns the matrix $\mathbf{A}\in \mathbb{C}^{N \times N}$ with $\mathbf{a}$ on the diagonal. The operator $ \otimes$ denotes the Kronecker product. The Euclidian norm is denoted by $\left\|\cdot\right\| $.

\vspace*{-1mm}
\section{System Model with IRS supported transmission}

 We consider communication from an $M$-antenna BS to a single-antenna user equipment (UE) as shown in Fig. 1. A planar IRS with $N$ elements (composed of $N_H$ horizontal and $N_V$ vertical) is located in between to assist. The locations of the BS and IRS are fixed whereas the UE can be in different locations.  Each element of the IRS has the ability to introduce a phase shift to an incoming narrowband signal. The phase is adjusted by an IRS-controller that enables manipulation of the impinging wave. The IRS-controller is connected to the BS over a backhaul link to coordinate between the IRS and BS. To configure the IRS elements, the CSI is crucial. Since the IRS is not equipped with radio frequency chains, we assume that the channel estimation is performed at the BS side.

\begin{figure}[h]
	\vspace*{-10mm}
	\flushleft
	\includegraphics[scale=0.34]{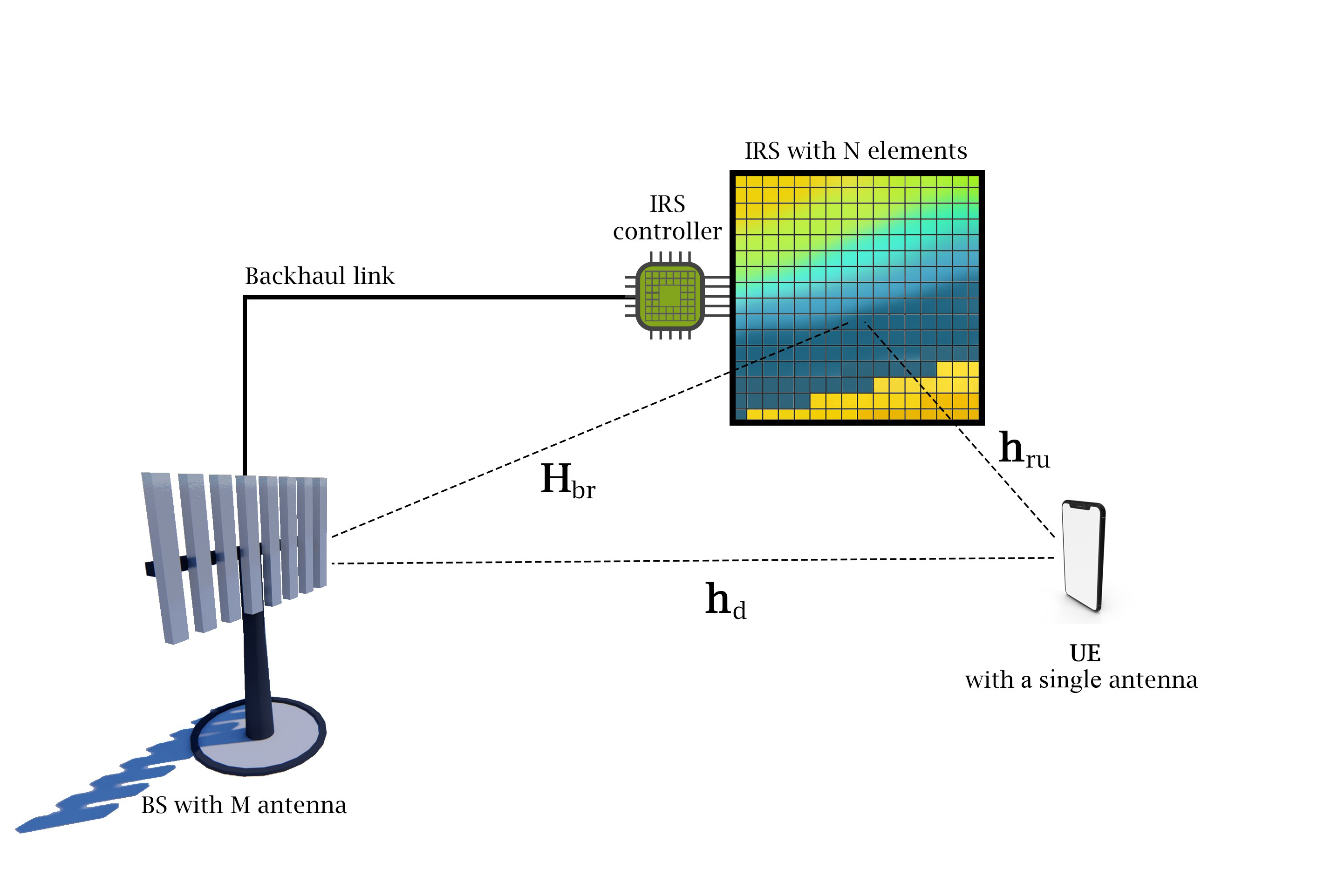}
		\vspace*{-5mm}
	\caption{Illustration of an IRS-assisted communication system. }\label{systemIllustration}
	\vspace*{-6mm}
\end{figure}

\vspace*{-1mm}
\subsection{Channel Estimation}

 We assume quasi-static flat-fading channels and the system operates in time divison duplex (TDD) mode. Pilot-based channel training is utilized to estimate the channels at the BS. During the channel estimation phase, the UE sends the pilot signal $x_t \in \mathbb{C}$ at time slot $t$. The received pilot signal at the BS is modeled as \cite{Jensen2019}
\begin{equation}\label{eq1}
 \mathbf{y}_t = \left( \mathbf{h}_\mathrm{d} + \mathbf{H}_\mathrm{br}\mathrm{diag}(\boldsymbol{\phi}_t)\mathbf{h}_\mathrm{ru}\right) x_t + \mathbf{n}_t,
\end{equation}
where $\mathbf{n}_t \sim \mathcal{CN}\left( \boldsymbol{0}, \mathbf{I}_M \right) $ is the additive white Gaussian noise (AWGN), $\mathbf{h}_\mathrm{d}  \in \mathbb{C}^{M\times 1}, \mathbf{H}_\mathrm{br} \in \mathbb{C}^{M \times N} $, $\mathbf{h}_\mathrm{ru} \in \mathbb{C}^{N\times 1}$ are the channels between BS and UE, BS and IRS, IRS and UE, respectively. The phase configuration at the IRS at time slot $t$ is denoted by $\boldsymbol{\phi}_t = [e^{j\phi_{t,1}}, \dots,e^{j\phi_{t,N}} ]^T \in \mathbb{C}^{N \times 1}$ where $\phi_{t,n} \in [0, 2\pi)$ is the phase shift of the $n$th element. 

We assume that the BS is equipped with a horizontal uniform linear array (ULA) placed on the $x$-axis. Unlike the UE, the IRS and BS have typically fixed locations once they are deployed. Therefore, $\mathbf{H}_\mathrm{br}$ is represented by a static line-of-sight (LoS) channel as $\mathbf{H}_\mathrm{br} = \sqrt{\beta_\mathrm{br}} \mathbf{a}_\mathrm{BS}(\varphi_\mathrm{BS},\theta_\mathrm{BS})\mathbf{a}_\mathrm{IRS}(\varphi_\mathrm{IRS}, \theta_\mathrm{IRS})^H$ where $\beta_\mathrm{br}$ is the pathloss coefficient, 
\begin{equation}\label{eqULA}
\mathbf{a}_\mathrm{BS}(\varphi_\mathrm{BS},\theta_\mathrm{BS}) = \left[ 1, \dots, e^{j2\pi (M-1)d_H \cos( \varphi_\mathrm{BS}) \cos( \theta_\mathrm{BS})}\right] ^T
\end{equation}
is the BS's array response vector where $\varphi_\mathrm{BS}$, $\theta_\mathrm{BS}$ are the azimuth and elevation angle-of-arrivals (AoA) to the IRS seen from the BS, $d_H$ is the antenna spacing parameter measured in the number of wavelengths.  The array response of the IRS (placed on the $yz$-plane) is denoted by 
\begin{equation}\label{eqUPA}
\mathbf{a}_\mathrm{IRS}(\varphi_\mathrm{IRS}, \theta_\mathrm{IRS}) = [e^{j\mathbf{k}(\varphi_\mathrm{IRS}, \theta_\mathrm{IRS})^T \mathbf{u}_1}, \dots, e^{j\mathbf{k}(\varphi_\mathrm{IRS}, \theta_\mathrm{IRS})^T \mathbf{u}_N}   ]^T 
\end{equation}
where $\varphi_\mathrm{IRS}$ and  $\theta_\mathrm{IRS}$ are the azimuth and elevation angle-of-departures (AoD) to the BS seen from the IRS, respectively. Recall that we consider a planar IRS. The wave vector is
\begin{equation}
\mathbf{k}(\varphi_\mathrm{IRS}, \theta_\mathrm{IRS}) = \frac{2\pi}{\lambda_c}\begin{bmatrix}
\cos( \varphi_\mathrm{IRS})\cos(\theta_\mathrm{IRS}) \\
\sin( \varphi_\mathrm{IRS})\cos(\theta_\mathrm{IRS}) \\
\sin(\theta_\mathrm{IRS})
\end{bmatrix}, 
\end{equation}
and the indexing vector is $\mathbf{u}_n = [0, i(n)d_r \lambda_c, j(n) d_r \lambda_c]^T$ where $\lambda_c$ is the wavelength at the carrier frequency, $i(n) = \mod(n-1, N_H),$ and $j(n) =  \lfloor (n-1)/N_H \rfloor $
are used for the describing the location of each IRS element \cite[Sec. 7.3]{massivemimobook}. The parameter $d_r$ denotes the element spacing at the IRS, in both the horizontal and vertical directions. Notice that the ULA array response in \eqref{eqULA} is a special case of planar array response in \eqref{eqUPA} where $[\mathbf{a}_\mathrm{BS}(\varphi_\mathrm{BS}, \theta_\mathrm{BS})]_m =  e^{j\mathbf{k}(\varphi_\mathrm{BS}, \theta_\mathrm{BS})^T \mathbf{u}_m}$ with  $\mathbf{u}_m = [ (m-1)d_H \lambda_c, 0,0]^T$.   

To account for the assumed limited scattering environment, the channels $\mathbf{h}_\mathrm{d}$ and $\mathbf{h}_\mathrm{ru}$ are represented by the Saleh-Valenzuela (SV) model \cite{Elbir2020, Ayach2014}. We assume that  there are  $L_\mathrm{d}$ and $L_\mathrm{ru}$ paths, respectively. Thus, the direct channel is modeled as
\begin{equation}	
 \mathbf{h}_\mathrm{d} =  \sqrt{\frac{1 }{L_\mathrm{d}}}\sum_{l=1}^{L_\mathrm{d}}  \alpha^l_\mathrm{d} \mathbf{a}_\mathrm{BS}(\varphi^l_\mathrm{BS},\theta^l_\mathrm{BS})
\end{equation}
where $\alpha^l_\mathrm{d} $ is the complex channel gain, $\varphi^l_\mathrm{BS}, \theta^l_\mathrm{BS} $ are the azimuth and elevation AoAs associated with the $l$th path.   Similarly, the channel between the IRS and UE is
\begin{equation}
 \mathbf{h}_\mathrm{ru} =  \sqrt{\frac{1}{L_\mathrm{ru}} }\sum_{l=1}^{L_\mathrm{ru}}  \alpha^l_\mathrm{ru} \mathbf{a}_\mathrm{IRS}(\varphi^l_\mathrm{IRS}, \theta^l_\mathrm{IRS} )
\end{equation}
where $\alpha^l_\mathrm{ru}$ is the complex channel gain, $\varphi^l_\mathrm{IRS}, \theta^l_\mathrm{IRS} $ are the azimuth and elevation  AoAs  associated with the $l$th path.

At time slot $t$, we can rewrite \eqref{eq1} as
\begin{equation}
\mathbf{y}_t = \left( \mathbf{h}_\mathrm{d} + \mathbf{V} \boldsymbol{\phi}_t\right) x_t + \mathbf{n}_t
\end{equation}
where   $\mathbf{V} = \mathbf{H}_\mathrm{br}\mathrm{diag}(\mathbf{h}_\mathrm{ru})= [\mathbf{v}_1,\mathbf{v}_2,\dots,\mathbf{v}_N] \in \mathbb{C}^{M \times N}$ is the cascaded BS-IRS-UE channel. The pilot signals are sent $T $ times by the UE. We assume that the channels are fixed during the estimation period and $\boldsymbol{\phi}_t$ is reconfigured at each time slot $t$.  The collection of all the pilot signal at the BS is $\mathbf{y}_p =[\mathbf{y}^T_1,\mathbf{y}^T_2, \dots,\mathbf{y}^T_T]^T \in \mathbb{C}^{TM\times 1}$ can be written as 
\begin{equation}
\mathbf{y}_p = \mathbf{X}\left( \boldsymbol{\Phi} \otimes \mathbf{I}_M \right)\mathbf{h} +\mathbf{n} 
\end{equation}
where the pilot signal is $\mathbf{X} = \mathrm{diag}\left( [x_1\mathbf{1}_M, \dots, x_T\mathbf{1}_M]\right) \in \mathbb{C}^{TM \times TM} $, and $\mathbf{n} \sim \mathcal{CN}\left(\boldsymbol{0}, \mathbf{I}_{TM} \right)$. The channels are stacked into  $\mathbf{h} = [\mathbf{h}^T_\mathrm{d}, \mathbf{v}^T_1, \dots, \mathbf{v}^T_N]^T  \in \mathbb{C}^{(N+1)M \times 1} $. All the phase configurations at the IRS are collected in $\boldsymbol{\Phi} = [ \boldsymbol{\bar{\phi}}_1,\dots,\boldsymbol{\bar{\phi}}_T]^T \in \mathbb{C}^{T \times (N+1)} $   where $\boldsymbol{\bar{\phi}}_t =[1 ,\boldsymbol{\phi}^T_t]^T  \in \mathbb{C}^{(N+1) \times 1} $ is the extended reflection pattern accounting for both the direct and cascaded channels. Notice that the first column of $\boldsymbol{\Phi}$ is set to an all one vector to estimate the direct channel.  

The IRS phase configuration  during the channel estimation period, $\boldsymbol{\Phi}$, mimics a discrete Fourier Transform matrix as in \cite{Jensen2019, You2020}. More precisely, each element of the phase matrix can be written as
\begin{equation}
\left[ \boldsymbol{\Phi}\right]_{t,n} = e^{-j\frac{ 2\pi (t-1)(n-1)}{N+1}}
\end{equation}
where $\boldsymbol{\Phi}$ can not contain more than $N+1$ unique values around the unit circle. Note that this specific selection of $\boldsymbol{\Phi}$ guarantess that  $\mathrm{rank}\left( \boldsymbol{\Phi}\right) = \min\left\lbrace T, N+1\right\rbrace $ and the phase of each element satisfies the unit-modulus constraint. Besides, the first column of $\boldsymbol{\Phi}$ is equal to an all one vector. The property $| [ \boldsymbol{\Phi}]_{t,n}|= 1$ is particulary important
since implementing different amplitudes at each IRS element can be costlier and harder. Another potential choice of $\boldsymbol{\Phi}$ that satisfies the same constraints is a truncated Hadamard matrix \cite{You2020}.

 Assuming that $T \geq N+ 1$, based on the pilot signal  $\mathbf{y}_p$, the channels can be estimated by the LS estimator  as  \cite{Jensen2019}
\begin{equation}\label{eqLS}
\hat{\mathbf{h}} = \arg \min_{\mathbf{h}} \left\| \mathbf{P} \mathbf{h}- \mathbf{y}_p\right\|^2_2 = \left(\mathbf{P}^H \mathbf{P}  \right)^{-1}\mathbf{P}^H \mathbf{y}_p
\end{equation}
where $\mathbf{P}  = \mathbf{X}\left( \boldsymbol{\Phi} \otimes \mathbf{I}_M \right)$ is the observation matrix. The BS can utilize these channel estimates to compute the downlink transmit beamforming vector at the BS and the optimum phase configuration at the IRS. Then, the BS can send the $N$ optimum phases to the IRS via backhaul link.

\subsection{IRS Phase Reconfiguration and Downlink Spectral Efficiency}

If the BS has perfect CSI, it  can compute the optimal phases and the beamforming vector  using the alternating optimization method in \cite{Wu2018a} as
\begin{equation}
\phi^\mathrm{opt}_n = \mathrm{arg}\left( \mathbf{h}_\mathrm{d}^H
\mathbf{w}  \right) - \mathrm{arg}\left( \mathbf{v}^H_n \mathbf{w}  \right), 
\end{equation}
\begin{equation}\label{eqBF}
\mathbf{w}^\mathrm{opt} = \frac{ \mathbf{h}_\mathrm{d} + \mathbf{V}\left( \boldsymbol{\phi}^\mathrm{opt}\right)^*  }{\left\|  \mathbf{h}_\mathrm{d} + \mathbf{V} \left(\boldsymbol{\phi}^\mathrm{opt}\right)^*  \right\| }
\end{equation}
where $\boldsymbol{\phi}^\mathrm{opt} = [\phi^\mathrm{opt}_1, \dots,\phi^\mathrm{opt}_N]^T \in \mathbb{C}^{N \times 1} $. We initialize the beamforming vector as  $ \mathbf{w}= \frac{1}{\sqrt{M}}[1,\dots,1]^T$. Note that the optimized phases are obtained by phase aligning the direct and cascaded channels. Besides, for any given phase configuration, the optimum transmit beamforming is equal to the maximum ratio precoding vector. 

 During the downlink transmission, the UE receives
\begin{equation}\label{eq2}
y_r =\left( \mathbf{h}^H_\mathrm{d} + \mathbf{h}^H_\mathrm{ru}\mathrm{diag}\left(  \boldsymbol{\phi}^\mathrm{opt}\right) \mathbf{H}^H_\mathrm{br}\right) \mathbf{w}^\mathrm{opt} s + n
\end{equation}
where $s$ is the data signal and $n \sim \mathcal{CN}(0,1)$ is the additive noise. Alternatively, we can rewrite \eqref{eq2} as
\begin{equation}
y_r =\left( \mathbf{h}^H_\mathrm{d} + \left(  \boldsymbol{\phi}^\mathrm{opt}\right)^T \mathbf{V}^H\right) \mathbf{w}^\mathrm{opt} s + n.
\end{equation}

If the channels are fixed throughout the transmission, the rate is 
\begin{align}
	R &= \log_2\left( 1 + \gamma \left| \left( \mathbf{h}^H_\mathrm{d} + \left(  \boldsymbol{\phi}^\mathrm{opt}\right)^T \mathbf{V}^H\right) \mathbf{w}^\mathrm{opt}\right|^2 \right) \label{eqSE} \\
	&=  \log_2\left( 1 + \gamma \left\|  \left( \mathbf{h}^H_\mathrm{d} + \left(  \boldsymbol{\phi}^\mathrm{opt}\right)^T \mathbf{V}^H\right) \right\|^2 \right)
\end{align}
where $\gamma$ is the signal-to-noise-ratio (SNR). If the BS utilizes the LS estimator then it treats the estimated channels as the true channels and calculates $\boldsymbol{\phi}^\mathrm{opt} $ and $\mathbf{w}^\mathrm{opt} $ based on $\hat{\mathbf{h}}$ in \eqref{eqLS}. Then, the optimum phase configuration  $\boldsymbol{\phi}^\mathrm{opt}$  based on LS estimator are sent to the IRS over the backhaul link.

\section{Deep Learning-based Phase Configuration}

According to the universal approximation theorem, a DNN has the capability of approximating any continuous function \cite{Goodfellow2016}. In supervised learning, DNNs are trained using a training dataset that is given as input-output pairs. The goal of the proposed DNNs is to  find the mapping between the received pilot signals and the optimum phase configuration and downlink transmit beamforming vector. The pilot signals go through all the channels and reach the BS. Therefore, it captures important information for the phase and beamforming setting since there is a nonlinear relation between the optimal phases and the channel coefficients. A properly designed DNN can learn this relation. Therefore, the problem is to train effectively the weights and biases of the DNN so that it can learn a nearly optimal mapping between received pilots and phases. A test dataset that is separately generated from the training data is used to evaluate the performance of the DNNs. During the online phase, the trained DNNs compute the required phases and beamforming vector.

As mentioned earlier, a main challenge of channel acquisition with IRS is that the number of channel coefficients increases proportionally to $N$. The conventional methods such as the LS estimator in \eqref{eqLS} requires a pilot training period with $T \geq N+1$. When applying an LS estimator and then treating the estimate as perfect, there is an information loss, which is not the case when we directly obtain the phase shifts and beamforming vector. Besides, the LS estimator is unaware of the underlying propagation conditions, while a DNN can learn it. Hence, it is possible for a DNN to outperform the conventional LS method. In this paper, we present two different DNNs with different $T$ values as described in the following subsections.

\vspace*{-2mm}

\subsection{Deep Learning Method 1}
In the first method, to train the DNN, we  set $T =N+1$  and use the input-output pairs  $\left\lbrace \mathbf{y}_p,  \boldsymbol{\Omega}\right\rbrace $ that are generated during the preamble stage. The output is formed by stacking the optimum phases and beamforming vector into $\boldsymbol{\Omega} = \left[ (\boldsymbol{\phi}^\mathrm{opt})^T,  (\mathbf{w}^\mathrm{opt})^T\right]^T \in \mathbb{C}^{(N + M) \times 1}$. Both input and output vectors contain complex numbers. To feed them into the DNN, the real and imaginary parts of each entry are separated. Thus, the input has size $2TM \times 1$ and the output dimension is $2(N+M) \times 1$. Using a training set of $n_\mathrm{train}$ samples consisting of different realizations, the DNN emulates the mapping by adjusting the  weights and bias terms. 

The proposed DNN (DL method 1) is composed of 3 fully connected hidden layers. The details are presented in Table I.  The input data is scaled using Standard Scaler function in the Python environment, which removes the mean and normalize the input data such that it has unit variance. We use the Adam optimizer with adaptive learning rates starting from $0.0005$. The learning rate is reduced to its half when there is no improvement in the last 5 epochs. As loss function, we select the mean square error (MSE). The batch size is chosen as $32$ and an early stopping criteria is applied that stops the training when the validation accuracy does not improve in 10 consecutive epochs. The maximum number of epochs is set to 200.

\begin{table}[]
	\centering
	\begin{tabular}{lllll}
		\multicolumn{1}{l|}{Layers} & \multicolumn{1}{l|}{Size} & Activation Function &  &  \\ \cline{1-3}
		\multicolumn{1}{l|}{Input} & \multicolumn{1}{l|}{$2TM$} &elu  &   \\
		\multicolumn{1}{l|}{Layer $1$ (Dense)} & \multicolumn{1}{l|}{$512$} & elu &   \\
		\multicolumn{1}{l|}{Layer $2$ (Dense)} & \multicolumn{1}{l|}{$512$} & elu &   \\
		\multicolumn{1}{l|}{Layer $3$ (Dense)} & \multicolumn{1}{l|}{$256$} & elu &    \\
		\multicolumn{1}{l|}{Output} & \multicolumn{1}{l|}{$2(N+M)$} & linear &    \\
		&                       &  &  
	\end{tabular}
\caption{Layout of the proposed DL method 1 where $T = N+1 $. }
	\vspace*{-4mm}
\end{table}
\vspace*{-2mm}
\subsection{Deep Learning Method 2}
In the second DNN, we set $T < N + 1$ to reduce the pilot overhead and the intention is that the DNN will learn how to reconstruct the channel despite the reduced dimensionality. The input-output pairs  $\left\lbrace \mathbf{y}_p, \boldsymbol{\Omega}\right\rbrace $ are generated during the preamble stage. Note that the input $\mathbf{y}_p$ is shorter in this case. As in  DL method 1, the real and imaginary parts of the complex signal are separated and then fed to the DNN.  DL method 2 consists of 4 fully connected hidden layers as presented in Table II. We use the same input scaling, batch size, Adam optimizer,  and loss function as in DL method 1.

\begin{table}[]
	\centering
	\begin{tabular}{lllll}
		\multicolumn{1}{l|}{Layers} & \multicolumn{1}{l|}{Size} & Activation Function &  &  \\ \cline{1-3}
		\multicolumn{1}{l|}{Input} & \multicolumn{1}{l|}{$2TM$} &elu  &   \\
		\multicolumn{1}{l|}{Layer $1$ (Dense)} & \multicolumn{1}{l|}{$500$} & elu &   \\
		\multicolumn{1}{l|}{Layer $2$ (Dense)} & \multicolumn{1}{l|}{$400$} & elu &   \\
		\multicolumn{1}{l|}{Layer $3$ (Dense)} & \multicolumn{1}{l|}{$400$} & elu &    \\
		\multicolumn{1}{l|}{Layer $4$ (Dense)} & \multicolumn{1}{l|}{$300$} & elu &    \\
		\multicolumn{1}{l|}{Output} & \multicolumn{1}{l|}{$2(N+M)$} & linear &    \\
		&                       &  &  
	\end{tabular}
	\caption{Layout of the proposed DL method 2 where $T < N+1 $.}
		\vspace*{-7mm}
\end{table}

\section{Numerical Results}

In this section, we evaluate the performance of the proposed DNNs  where $M=10$ and $N=100$. For each data sample, the location of the UE with height $1.5$ m is drawn from a uniform distribution over a $10 \times 10$ square-meter room. The numbers of paths are set as $L_\mathrm{d} = L_\mathrm{ru} = 5$.  The downlink transmit power is $10$ dBm and  the pilot power is $25$ dBm, unless otherwise stated. The receiver noise power is $-94$ dBm where the bandwidth is $20$ MHz.
 
The pathloss coefficient of the BS-IRS channel is calculated as $\beta_\mathrm{br} = \frac{ N A}{4 \pi d^2_\mathrm{br}}$ where $A = (d_r \lambda_c)^2$ is the  area of one IRS element with $d_r = 0.25$ and $ \lambda_c = 0.1$ m and $d_\mathrm{br} = 292$ m is the distance between the BS and IRS.  The antenna spacing at the BS is $d_H = 0.5$.

The other pathloss parameters are set based on \cite{Saleh1987, Wang2020} as
$\alpha^l_\mathrm{d} = \sqrt{\beta_0 (d_\mathrm{bu} /d_0)^{-3.8}} e^{-j2\pi f_c \tau^l_\mathrm{d} }$ and $ \alpha^l_\mathrm{ru} = \sqrt{\beta_0 (d_\mathrm{ru}/d_0)^{-3.8}} e^{-j2\pi f_c \tau^l_\mathrm{ru} }$
where $d_0 = 1$ m, $\beta_0 = -20.4$ dB is the reference pathloss, $d_\mathrm{bu}$ and $d_\mathrm{ru}$  are the distances between BS-UE and IRS-UE, respectively.   The associated path delays in nanoseconds are $\tau^l_\mathrm{d}\sim \mathcal{U}\left[0, 10 \right]$, $\tau^l_\mathrm{ru}\sim \mathcal{U}\left[0, 5  \right] $. The minimum allowed  $d_\mathrm{ru} = 7$ m.

The DNN was trained based on a dataset of $n_\mathrm{train} = 80000$ training samples. Particularly, $80\%$ of the samples was used for training and $20\%$ for  validation. Another $2000$ samples formed the test dataset, which is independent from the training dataset but drawn from the same distribution. The training process takes around 1 hour and the online testing  requires approximately 0.2 ms for both methods in Python on a Windows 10 personal computer having  Intel i7-6600U CPU with 2.81 GHz and Intel HD Graphics 520 GPU.

The normalized mean-squared-error (NMSE) of the phase configuration is calculated as 
\begin{equation}\label{nmse}
	\mathrm{NMSE} = \frac{1}{n_\mathrm{test}} \sum_{s=1}^{n_\mathrm{test}} \frac{\left\| \boldsymbol{\phi}^\mathrm{opt}_s - \boldsymbol{\hat{\phi}}^x_s \right\|^2 }{\left\|\boldsymbol{\phi}^\mathrm{opt}_s  \right\|^2}
\end{equation}
where $ \boldsymbol{\phi}^\mathrm{opt}_s$ is the optimum phase configuration based on perfect CSI, $\boldsymbol{\hat{\phi}}^x_s $ is either the output of one of the DNNs or calculated based on LS-based estimation i.e., $x \in \left\lbrace \text{DL method 1}, \text{DL method 2}, \text{LS-based method}\right\rbrace $. Notice that $\left\|\boldsymbol{\phi}^\mathrm{opt}_s  \right\|^2 = \| \boldsymbol{\hat{\phi}}^x_s \|^2 =N$.

\begin{figure}[h]
\vspace*{-4mm}
	\includegraphics[scale=0.10]{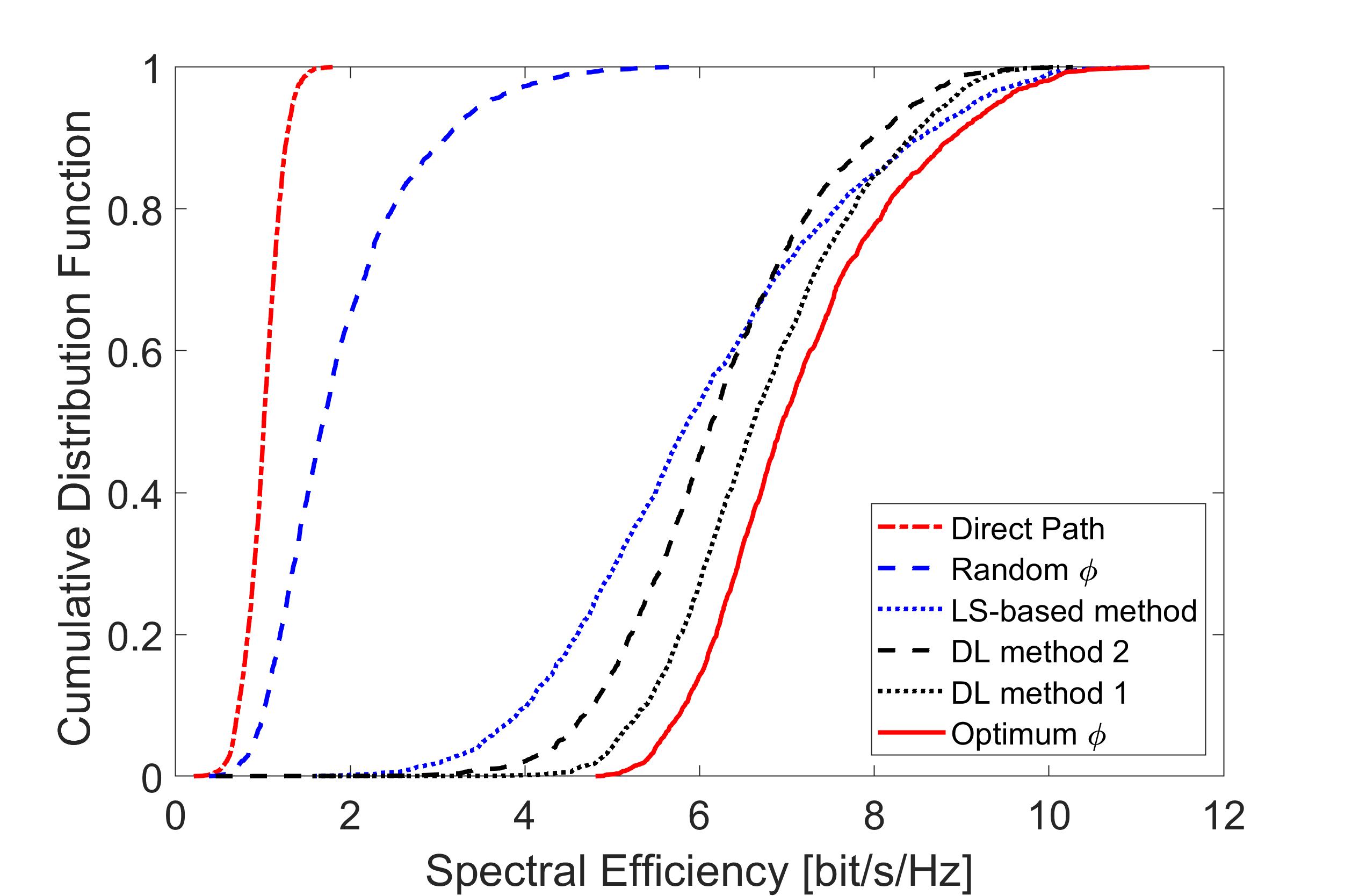}
	\caption{Cumulative distribution function of the downlink spectral efficiency.}\label{fig1}
	\vspace*{-4mm}
\end{figure}

\begin{figure}[h]
	\vspace*{-2mm}
	\includegraphics[scale=0.10]{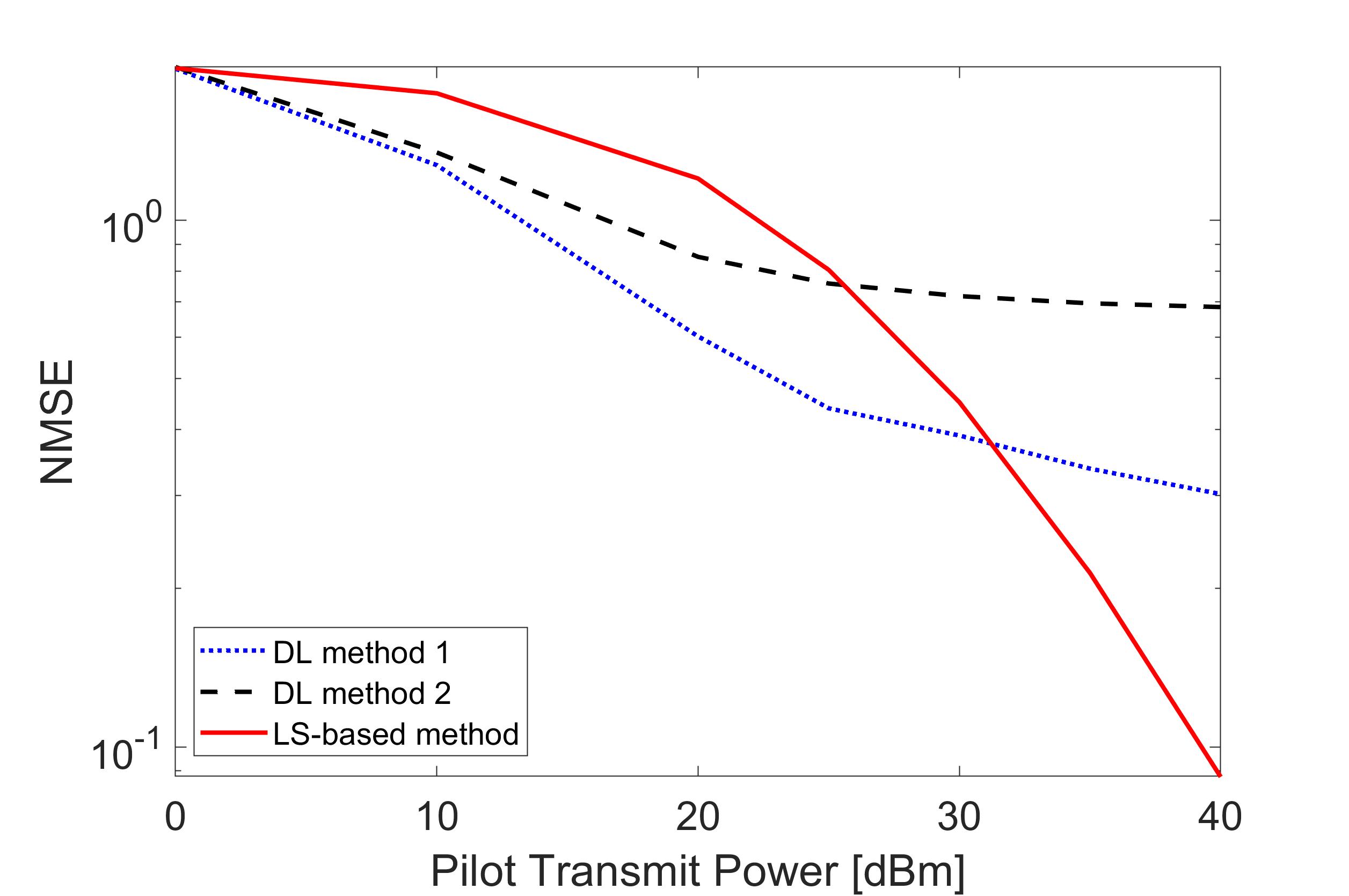}
	\caption{NMSE versus  pilot transmit powers.}\label{fig3}
	\vspace*{-3mm}
\end{figure}

\begin{figure}[h]
	\vspace*{-3mm}
	\includegraphics[scale=0.10]{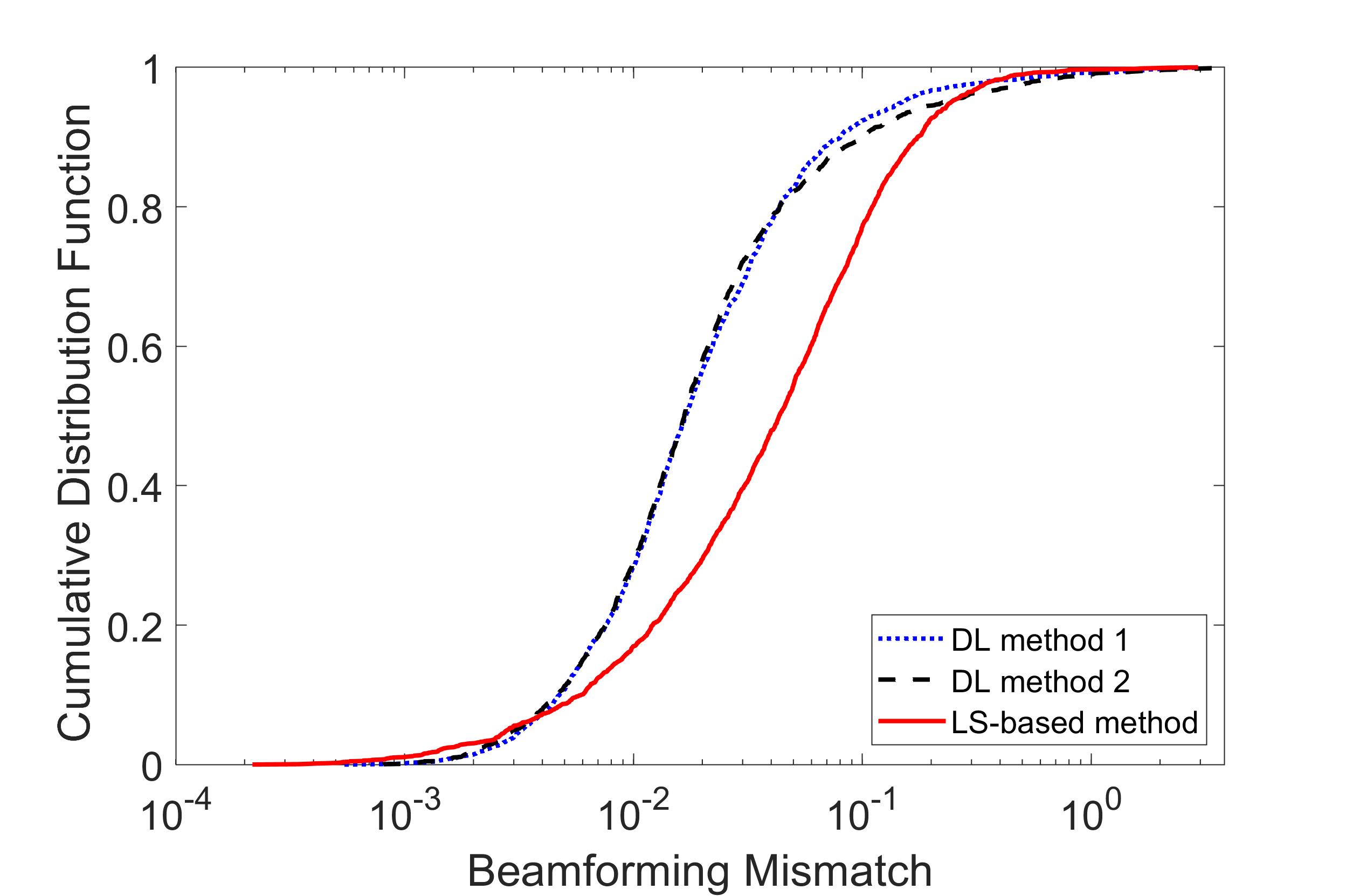}
	\caption{Cumulative distribution function of beamforming mismatch for different methods.}\label{fig4}
	
\end{figure}

Fig. \ref{fig1} compares the cumulative distribution of the downlink spectral efficiencies that are calculated based on \eqref{eqSE} for different cases.  The ``Direct Path'' label represents the case when there is no IRS in the system.  The  ``Random $\boldsymbol{\phi}$''   denotes the setting where the phase configuration at the IRS is set randomly and  the downlink transmit beamforming vector is calculated based on these phases for each test sample.  We observe that DL method 1 performs better than the classical LS-based method for almost all of the samples. It is very close to the ``Optimum $\boldsymbol{\phi}$'' in which the phase configuration  and the beamforming vector are computed based on perfect CSI. Note that in both DL method 1 and the LS-based method, we used the same pilot length $T = N + 1 = 101$. Moreover,  DL method 2 in which we used $T=64$ also performs better than the LS-based method for most of the test data. The pilot overhead is reduced by $35\%$ in  DL method 2 compared to  DL method 1 and LS-based method. This is  because of the fact that the DNNs are able to find the direct mapping between the received pilot signals and the optimum phases and beamformer whereas the LS-based method treats the estimates as the true channels that causes an information loss. Besides, the LS estimator does not have any prior information on the channel whereas the DNNs can learn the features of the channel from the datasets. 

In Fig. \ref{fig3}, we compare the NMSEs of the presented methods for different pilot transmit powers. During the preamble stage, the training data is generated for different pilot transmit powers while keeping the other parameters fixed. Then, the DNNs are trained by these received pilots. It is demonstrated that for practical pilot powers the DL methods provide better performance whereas for high pilot powers the LS-based method outperforms the DL approaches.  However, potentially, another DNN could be designed and trained for high pilot powers by increasing the width of the hidden layers that would increase the accuracy. However, a potential pitfall with this approach is to create an overfitting problem causing the DNN to memorize the training set.

In Fig. \ref{fig4}, we compare the accuracy of the downlink transmit beamforming vectors that are designed at the BS side based on the presented methods. More precisely, the beamforming mismatch is computed as $ \left\| \mathbf{w}^\mathrm{opt} - \mathbf{w}^x \right\|^2 $ where $x \in \left\lbrace \text{DL method 1}, \text{DL method 2}, \text{LS-based method}\right\rbrace $. Notice that $\| \mathbf{w}^\mathrm{opt} \| = \| \mathbf{w}^x \| = 1$. We observe that the DL methods give very similar accuracy and they are superior to the LS-based approach.

\section{Conclusions}
This paper proposes a DNN framework  for the reconfiguration of IRS elements based on the available pilot signals. We showed that a properly trained feed-forward DNN is able to learn how to configure the IRS phases  and downlink beamforming vector.  DL method 1 outperforms the classical LS estimator based method for practical pilot transmit powers. Its performance is close to the perfect CSI based approach. In addition,  DL method 2 reduces the pilot overhead and have a similar performance to the LS based method.

To further improve the framework, other things could be done such as considering multiple users, IRS-element grouping for reducing the pilot overhead further or using quantized IRS phases. Besides, measured channels could be used for DNN training.



\bibliographystyle{IEEEtran}
\bibliography{IEEEabrv,refs}
%

\end{document}